\documentclass[twocolumn,showpacs,showkeys]{revtex4}
\usepackage{graphicx}

\newcommand{\bastar}{\begin{eqnarray*}}
\newcommand{\eastar}{\end{eqnarray*}}
\newskip\humongous \humongous=0pt plus 1000pt minus 1000pt

\relax
\newcommand{\be}{\begin{equation}}
\newcommand{\ee}{\end{equation}}
\newcommand{\bea}{\begin{eqnarray}}
\newcommand{\eea}{\end{eqnarray}}
\newcommand{\X}{{\vec X}}
\newcommand{\pro}{\partial}
\newcommand{\n}{\hat n}
\newcommand{\oneg}{\displaystyle\frac{1}{g}}

\newcommand{\D}{{\hat D}}

\newcommand{\A}{{\vec A}}
\newcommand{\valpha}{{\vec \alpha}}

\newcommand{\dfrac}{\displaystyle\frac}
\newcommand{\ba}{\begin{array}}
\newcommand{\ea}{\end{array}}

\newcommand{\nn}{\nonumber}
\newcommand{\hn}{\hat n}
\begin{document}
\title{Chromoelectric Knot in QCD}
\author{Y. M. Cho}
\email{ymcho@yongmin.snu.ac.kr}
\affiliation{C. N. Yang Institute
for Theoretical Physics, \\
State University of New York, Stony Brook, New York 11794, USA \\
and \\
School of Physics, College of Natural Sciences, Seoul National University,
Seoul 151-742, Korea\\}
\begin{abstract}
~~~~~We argue that the Skyrme theory describes the chromomagnetic 
(not chromoelectric) dynamics of QCD. This shows that 
the Skyrme theory could more properly be interpreted 
as an effective theory which is dual to QCD, rather than an
effective theory of QCD itself. This leads us to predict the existence of a 
new type of topological knot, a twisted chromoelectric flux ring, 
in QCD which is dual to the chromomagnetic Faddeev-Niemi knot 
in Skyrme theory. We estimate the mass and the decay width
of the lightest chromoelectric knot to be around $50~GeV$ and 
$117~MeV$. 
\end{abstract}
\maketitle

Recently Faddeev and Niemi have conjectured the existence of 
a topological knot in quantum chromodynamics (QCD), a twisted 
chromomagnetic vortex ring which is similar to the Faddeev-Niemi knot
in Skyrme theory \cite{fadd1,fadd2}. This is an interesting conjecture
based on the popular view that the Skyrme theory 
is an effective theory of strong interaction. 
{\it The purpose of this paper is to predict
the existence of a topological glueball in QCD
made of the twisted chromoelectric flux ring,
which is dual to Faddeev-Niemi knot in Skyrme theory.
We estimate the mass of the lightest knot glueball
to be around 50 $GeV$}. Although topological, the chromoelectric
knot could be cut and decay to lowlying hadrons, due to the presence
of the quarks and gluons in the theory.

The Skyrme theory has played an important role in physics,
in particular in nuclear physics as a successful effective
field theory of strong interaction \cite{skyr,prep,rho,witt}.
A remarkable feature of Skyrme theory is its rich
topological structure \cite{cho01}. It has been
known that the theory allows (not only the original skyrmion
but also) the baby skyrmion and the Faddeev-Niemi knot
\cite{piet,fadd2}. More importantly, it contains a (singular) monopole
which plays a fundamental role. In fact all the finite energy
topological objects in the theory could be viewed either as
dressed monopoles or as confined magnetic flux
of the monopole-antimonopole pair, confined by the Meissner effect. 
This observation has led us to propose that the theory can be
interpreted as a theory of monopoles, in which the
magnetic flux of the monopole-antimonopole pairs is confined
by the Meissner effect \cite{cho01}.

This implies that it should be interpreted as
an effective theory of strong interaction
which is dual to QCD, rather than
an effective theory of QCD itself. This is because in QCD it is
not the monopoles but the quarks which are confined.
And QCD confines the chromoelectric flux
with a dual Meissner effect.
This is in sharp contradiction with the popular view that 
the Skyrme theory is an effective theory of QCD. In
the following we compare the two contrasting views, and propose 
a simple experiment which can tell which view is the correct one.

Let $\omega$ and $\hat n$ (with ${\hat n}^2 = 1$) be the Skyrme 
field and the non-linear sigma field, and let
\bea
&U = \exp (\dfrac{\omega}{2i} \vec \sigma \cdot \hat n) 
= \cos \dfrac{\omega}{2} - i (\vec \sigma \cdot \hat n)
\sin \dfrac{\omega}{2}, \nn\\
&L_\mu = U\partial_\mu U^{\dagger}.
\label{su2}
\eea
With this one can write the Skyrme Lagrangian as \cite{skyr}
\bea
&{\cal L} = \dfrac{\mu^2}{4} {\rm tr} ~L_\mu^2
+ \dfrac{\alpha}{32}
{\rm tr} \left( \left[ L_\mu, L_\nu \right] \right)^2, 
\label{slag}
\eea
where $\mu$ and $\alpha$ are the coupling constants.
The Lagrangian has a hidden $U(1)$ gauge symmetry
as well as a global $SU(2)$ symmetry.
With the spherically symmetric ansatz 
and the boundary condition
\bea
&\omega = \omega (r),~~~~~\hat n = \hat r, \nn\\
&\omega(0)= 2\pi,~~~~~\omega(\infty)= 0,
\label{skans}
\eea
one has the well-known skyrmion which has a finite
energy $E \simeq 73~{\sqrt \alpha} \mu$ \cite{skyr}.
It carries the baryon number
\bea
&N_s = \dfrac{1}{8\pi^2} \int \epsilon_{ijk} N_{ij} (\partial_k \omega)
\sin^2 \dfrac{\omega}{2} d^3r =1, \nn\\
&N_{ij} = \hn \cdot (\partial_j \hn \times \partial_k \hn),
\label{bn}
\eea
which represents the non-trivial homotopy $\pi_3(S^3)$
described by $U$ in (\ref{su2}). It also carries the magnetic charge
\bea
N_m = \dfrac{1}{4\pi} \int \epsilon_{ijk} N_{ij} d\sigma_k
= 1,
\eea
which represents the homotopy $\pi_2(S^2)$ of the monopole
described by $\hn$ \cite{cho01}.

A remarkable point of the Skyrme theory is
that $\omega=\pi$ becomes a classical solution, independent 
of $\hn$. So restricting $\omega$ to $\pi$, one can
reduce the Skyrme Lagrangian (\ref{slag}) to
the Skyrme-Faddeev Lagrangian
\bea
{\cal L}_{SF}= -\dfrac{\mu^2}{2} (\partial_\mu \hat
n)^2-\dfrac{\alpha}{4}(\partial_\mu \hat n \times
\partial_\nu \hat n)^2,
\label{sflag}
\eea
whose equation of motion is given by
\bea
&\hn \times \partial^2 \hn
+ \dfrac{\alpha}{\mu^2} (\partial_\mu N_{\mu\nu})
\partial_\nu \hn = 0, \nn\\
&N_{\mu\nu} = \hn \cdot (\partial_\mu \hn \times \partial_\nu \hn)
=\partial_\mu C_\nu - \partial_\nu C_\mu.
\label{sfeq}
\eea
It is this equation that allows not only
the baby skyrmion and the Faddeev-Niemi
knot but also the non-Abelian monopole (Notice that $N_{\mu\nu}$
forms a closed two-form, so that it admits a potential 
at least locally sectionwise). 
This indicates that the Skyrme theory has a $U(1)$ gauge 
symmetry \cite{cho01}

With
\bea
\hat C_\mu = -\dfrac{1}{g} \hn \times \partial_\mu \hn,
\label{ccon}
\eea
the Lagrangian (\ref{sflag}) can be put into a very
suggestive form \cite{cho01,cho02},
\bea
&{\cal L}_{SF} = -\dfrac{\alpha}{4} \hat H_{\mu\nu}^2
- \dfrac{\mu^2} {2} \hat C_\mu^2, \nn\\
&\hat H_{\mu\nu} = \partial_\mu \hat C_\nu - \partial_\nu \hat C_\mu
+ g \hat C_\mu \times \hat C_\nu.
\label{qcdlag}
\eea
Actually with $\sigma=\cos~(\omega/2)$
the Skyrme Lagrangian (\ref{slag}) itself
can be expressed as 
\bea
&{\cal L} =-\dfrac{\alpha}{4} g^2 (1-\sigma^2)^2 \hat H_{\mu\nu}^2
-\dfrac{\mu^2}{2} g^2 (1-\sigma^2) \hat C_\mu^2 \nn\\
&-\dfrac{\mu^2}{2} \dfrac{(\partial_\mu \sigma)^2}{1-\sigma^2}
-\dfrac{\alpha}{4} g^2 (\partial_\mu
\sigma \hat C_\nu - \partial_\nu \sigma \hat C_\mu)^2 \nn\\
&\simeq -\dfrac{\alpha}{4} g^2 \hat H_{\mu\nu}^2 
-\dfrac{\mu^2}{2} g^2  \hat C_\mu^2 \nn\\
&-\dfrac{\mu^2}{2} (\partial_\mu \sigma)^2 
-\dfrac{\alpha}{4} g^2 (\partial_\mu
\sigma \hat C_\nu - \partial_\nu \sigma \hat C_\mu)^2. 
\label{slag1}
\eea
The approximation holds for 
small $\sigma$, which describes a linearized Skyrme theory. 
In this expression the Skyrme theory 
assumes the form of a massive
gauge theory (interacting with the scalar field $\sigma$)
in which the gauge potential is restricted by (\ref{ccon}).

To amplify this point further, consider the $SU(2)$ QCD for simplicity.
Introducing an isotriplet unit vector field $\n$ which selects
the color charge direction (i.e., the ``Abelian'' direction)
at each space-time point, we can
decompose the gauge potential into the restricted
potential $\hat B_\mu$
and the gauge covariant vector field $\vec X_\mu$ \cite{cho80,cho81},
\bea
& \vec{A}_\mu =A_\mu \n - \oneg \n\times\pro_\mu\n+\X_\mu\nonumber
= \hat B_\mu + \X_\mu, 
\label{cdec}
\eea
where $A_\mu = \n\cdot \vec A_\mu$
is the ``electric'' potential.
Notice that the restricted potential is precisely the connection which
leaves $\n$ invariant under the parallel transport,
\bea
\D_\mu \n = \pro_\mu \n + g {\hat B}_\mu \times \n = 0.
\eea
Under the infinitesimal gauge transformation
\bea
\delta \n = - \vec \alpha \times \n  \,,\,\,\,\,
\delta \A_\mu = \oneg  D_\mu \vec \alpha,
\eea
one has
\bea
&&\delta A_\mu = \oneg \n \cdot \pro_\mu \valpha,\,\,\,\
\delta \hat B_\mu = \oneg \D_\mu \valpha  ,  \nn \\
&&\hspace{1.2cm}\delta \X_\mu = - \valpha \times \X_\mu  .
\eea
This shows that $\hat B_\mu$ by itself describes
an $SU(2)$ connection which
enjoys the full $SU(2)$ gauge degrees of freedom. Furthermore
$\vec X_\mu$ transforms covariantly under the gauge transformation.
Most importantly, the decomposition (\ref{cdec}) is
gauge-independent. Once the color direction $\hn$ is selected the
decomposition follows automatically, independent of
the choice of a gauge.

The advantage of the decomposition (\ref{cdec})
is that all the topological features of the original
non-Abelian gauge theory are explicitly inscribed in $\hat B_\mu$.
The isolated singularities of $\hat{n}$ defines $\pi_2(S^2)$
which describes the Wu-Yang monopole \cite{cho80,cho81}.
Besides, with the $S^3$
compactification of $R^3$, $\hat{n}$ characterizes the
Hopf invariant $\pi_3(S^2)\simeq\pi_3(S^3)$ which describes
the topologically distinct vacua and the instantons \cite{cho02,cho79}.
The importance of the decomposition has recently been 
appreciated by many authors in studying various aspects 
of QCD \cite{fadd1,gies}. Furthermore in mathematics
the decomposition plays a crucial role in studying 
the geometrical aspects (in particular the Deligne cohomology)
of non-Abelian gauge theory \cite{cho75,zucc}.

Notice that the restricted potential $\hat{B}_\mu$
actually has a dual structure.
Indeed the field strength made of the restricted potential is decomposed as
\begin{eqnarray}
&\hat{B}_{\mu\nu}=\hat F_{\mu\nu}+ \hat H_{\mu\nu}
=(F_{\mu\nu}+H_{\mu\nu}) \hn, \nonumber\\
&F_{\mu\nu}=\partial_\mu A_{\nu}-\partial_{\nu}A_\mu, \nn\\
&H_{\mu\nu}=-\dfrac{1}{g} N_{\mu\nu} 
=-\dfrac{1}{g} (\partial_\mu C_\nu-\partial_\nu C_\mu),
\end{eqnarray}
where now $C_\mu$ plays the role of the ``magnetic'' 
potential \cite{cho80,cho81}.
This shows that the gauge potential (\ref{ccon}) which
appears in the Skyrme-Faddeev Lagrangian (\ref{qcdlag})
is precisely the chromomagnetic potential of QCD.

With (\ref{cdec}) we have
\bea
\vec F_{\mu\nu}=\hat B_{\mu \nu} + \D_\mu \X_\nu -
\D_\nu \X_\mu + g\X_\mu \times \X_\nu,
\eea
so that the Yang-Mills Lagrangian is expressed as
\bea
&{\cal L}_{QCD}=-\dfrac{1}{4}
{\hat B}_{\mu\nu}^2 -\dfrac{1}{4} ( \D_\mu \X_\nu -
\D_\nu \X_\mu)^2 \nn\\
&-\dfrac{g}{2} {\hat B}_{\mu\nu}
\cdot (\X_\mu \times \X_\nu)
- \dfrac{g^2}{4} (\X_\mu \times \X_\nu)^2.
\eea
This tells that QCD can be viewed
as a restricted gauge theory made of the binding gluon
$\hat B_\mu$, which has the valence gluon 
$\X_\mu$ as a gauge covariant colored source \cite{cho80,cho81}.
Now, suppose that the confinement mechanism generates
a mass $\mu$ for the binding gluon.
Then, in the absence of $A_\mu$
and $\vec X_\mu$, the above Lagrangian reduces
exactly to the Skyrme-Faddeev Lagrangian (\ref{sflag}).
Furthermore, with
\bea
&A_\mu = \partial_\mu \sigma,
~~~\X_\mu = f_1 \partial_\mu \n + f_2 \n \times \partial_\mu \n \nn\\
&\phi = f_1 + if_2,~~~~~\partial_\mu \phi = 0,
\eea
we have
\bea
&{\cal L}_{QCD} \simeq -\dfrac{(1-g \phi^* \phi)^2}{4} g^2 \hat H_{\mu\nu}^2
-\dfrac{\mu^2}{2} g^2 \hat C_\mu^2 \nn\\
&-\dfrac{\mu^2}{2} (\partial_\mu \sigma)^2
-\dfrac{\phi^* \phi}{4} g^2 (\partial_\mu \sigma \hat C_\mu
-\partial_\nu \sigma \hat C_\nu)^2.
\eea
This (with $\alpha=(1-g \phi^* \phi)^2=\phi^* \phi$) 
is precisely the linearized Skyrme Lagrangian
in (\ref{slag1}). So, if we like, we can actually derive the 
linearized Skyrme theory from QCD with simple assumptions \cite{cho01}.
This shows how the Skyrme theory stems from
QCD. More importantly, this reveals that the Skyrme theory
describes the chromomagnetic dynamics, not the chromoelectric
dynamics, of QCD.

Just like the $SU(2)$ QCD the Lagrangian (\ref{sflag}) has
the non-Abelian monopole solution \cite{cho01}. It also has 
a magnetic vortex solution known as the baby skyrmion and a twisted vortex 
solution known as the helical baby skyrmion \cite{cho01,piet}. 
The existence of the vortex solutions implies the existence of
the Meissner effect in Skyrme theory. To see how the
Meissner effect comes about, notice that due to the $U(1)$
gauge symmetry the theory has a conserved current
\bea
&j_\mu = \pro_\nu N_{\mu\nu},~~~~~\pro_\mu j_\mu = 0.
\label{sc}
\eea
Clearly this is the current which generates the Meissner effect
and confines the magnetic field of the vortex \cite{cho01}.
This confirms that the Skyrme theory indeed has a built-in
Meissner effect and confinement mechanism.

More importantly the Skyrme theory admits the Faddeev-Niemi knot,
which is nothing but the twisted magnetic vortex ring made of 
the helical baby skyrmion \cite{cho01}.
It has the knot quantum number \cite{cho01,fadd1}
\bea
N_k = \dfrac{1}{32\pi^2} \int \epsilon_{ijk} C_i N_{jk} d^3x=1.
\label{kqn}
\eea
Obviously the knot has a topological stability.
Furthermore, this topological stability is now
backed up by the dynamical stability. To see this, notice that
the chromoelectric supercurrent (\ref{sc}) has two components,
the one moving along the knot and
the other moving around the knot tube. And the
supercurrent moving along the knot generates an angular momentum
around the $z$-axis which provides the centrifugal force
preventing the vortex ring to collapse. Put it differently, the
supercurrent generates a magnetic flux trapped
in the knot disk which can not be squeezed out. And this
flux provides a stabilizing repulsive force which prevent the
collapse of the knot. This is how the knot acquires the dynamical
stability.

One could estimate the energy of the knot. Theoretically 
it has been shown that the knot energy has the following bound
\cite{ussr}
\bea
c~\sqrt{\alpha}~\mu~N^{3/4} \leq E_N
\leq C~\sqrt{\alpha}~\mu~N^{3/4},
\label{ke}
\eea
where $C$ is an unknown constant equal to or larger than $c$.
This suggests that the knot energy is proportional to $N^{3/4}$.
Indeed numerically, one finds \cite{batt}
\bea
E_N \simeq 252~\sqrt{\alpha}~\mu~N^{3/4},
\label{nke}
\eea
up to $N=8$. This sub-linear $N$-dependence of knot energy 
means that a knot with large $N$ can not decay
to the knots with smaller $N$.

Adopting the popular view that the Skyrme theory 
is an effective theory of QCD one can easily predict the existence
of a chromomagnetic knot in QCD. Furthermore one can estimate 
the mass of this knot from (\ref{nke}).
In this picture the parameters 
$\mu$ and $\alpha$ may be chosen to be \cite{prep,rho}
\bea
\mu=f_{\pi} \simeq 93~MeV,
~~~~~\alpha=8\epsilon^2 \simeq 0.0442,
\label{data1}
\eea
with the baryon mass $m_b \simeq 1.427~GeV$.
In a slightly different fitting one may choose \cite{prep,witt}
\bea
\mu=f_{\pi} \simeq 65~MeV,
~~~~~\alpha=8\epsilon^2 \simeq 0.0336.
\label{data2}
\eea
to have the baryon mass $m_b \simeq 0.870~GeV$. 
So from (\ref{data1}) we find
the mass of the lightest glueball to be
\bea
m_k \simeq 4.93~GeV,
\eea
but with (\ref{data2}) we obtain
\bea
m_k \simeq 3.00~GeV.
\eea
From this we expect the mass of the knot glueball proposed 
by Faddeev and Niemi to be around 3 to 5 $GeV$.

Our result in this paper challenges this traditional view.
We have shown that the Skyrme theory describes the chromomagnetic
(not chromoelectric) dynamics of QCD. Moreover,
the real baryon is made of quarks
which carry the chromoelectric charge, while
the skyrmion is actually a dressed monopole 
which carries the magnetic charge. And
the Faddeev-Niemi knot is made of the color magnetic flux,
while the glueball in QCD 
is supposed to carry the color electric flux.
Furthermore, although our analysis
implies that the Skyrme theory is a theory of confinement,
what is confined here is the monopoles, not the quarks.
And what confines the quarks in QCD is a dual Meissner effect, not
the Meissner effect. This tells that the Skyrme theory may
not be viewed as an effective theory of QCD, but more properly
as an effective theory which is dual to QCD.

This dual picture implies that QCD could admit
a chromoelectric knot which is dual to the chromomagnetic
Faddeev-Niemi knot. This is because one could make such a knot
by twisting a $g \bar g$
flux and smoothly connecting both ends.
Assuming the existence one may estimate the mass
of the knot. In this case one may identify
${\sqrt \alpha}\mu$ as the QCD scale $\Lambda_{QCD}$, 
because this is the only scale we have in QCD.
So, with \cite{pesk}
\bea
\Lambda_{QCD} \simeq {\sqrt \alpha}~\mu \simeq 200~MeV,
\label{sqcd}
\eea
one can easily estimate the mass of the lightest electric knot.
From (\ref{nke}) we expect
\bea
M_k \simeq 50~GeV.
\eea
The stability of such chromoelectric knot is far from guaranteed. This
is because in QCD we have other fields, the quarks and gluons,
which could destabilize the knot. For example, the knot can be cut
and decay to $g \bar g$ pairs and thus to
lowlying hadrons. We could estimate the decay width of the knot
from the one-loop effective action of QCD.
According to the effective action 
the chromoelectric background is unstable and decays
to $g \bar g$, with the probability
$11g^2 E^2/96\pi$ per unit volume per unit time \cite{cho1,cho2}.
So assuming that the knot is made of $g \bar g$ flux ring
of thickness $1/\Lambda_{QCD}$ and radius of about $3/\Lambda_{QCD}$,
we can estimate the decay width $\Gamma$ of the knot
\bea
&\Gamma \simeq \dfrac{11g^2}{96\pi} 
\Big(\dfrac{g\Lambda_{QCD}^2}{\pi} \Big)^2
\times \dfrac{6\pi^2}{\Lambda_{QCD}^3} 
\simeq 11\pi \alpha_s^2~\Lambda_{QCD} \nn\\ 
&\simeq 117 MeV, 
\label{dwk}
\eea
where we have put $\alpha_s(M_k) \simeq 0.13$ \cite{pesk}. Of course
this is a rough estimate, but this implies that the chromoelectric knot
can have a typical hadronic decay. In the presence of quarks, 
a similar knot made of a twisted $q\bar q$ flux
could also exist in QCD.

In this paper we have challenged the popular view of
Skyrme theory, and provided an alternative view.
There is a simple way to determine which is the correct view. 
This is because the two views predict totally different knot
glueballs which could be verified by the experiments. 
We have argued that the knot in traditional view
is a chromomagnetic knot, while the knot we predict here is a
chromoelectric knot. More importantly, we have shown 
that in the traditional view the mass of
the lightest knot glueball should be around 3 to 5 $GeV$, but in
the dual picture the mass of such glueball should be around 50
$GeV$. So, experimentally one could tell which is the correct view
simply by measuring the mass of the
exotic knot glueball. Certainly the LHC could be an ideal place
to determine which view is correct. The details of our argument 
will be published elsewhere \cite{cho3}.

{\bf ACKNOWLEDGEMENT}

~~~The author thanks Professor C. N. Yang for the
illuminating discussions, and G. Sterman for the kind hospitality
during his visit to Institute for Theoretical Physics. The work is
supported in part by the ABRL Program (Grant
R14-2003-012-01002-0) of Korea Science and Engineering 
Foundation, and by the BK21 project
of the Ministry of Education.

\end{document}